\def\year{2022}\relax
\documentclass[letterpaper]{article} 
\usepackage{aaai22}  
\usepackage{times}  
\usepackage{helvet}  
\usepackage{courier}  
\usepackage[hyphens]{url}  
\usepackage{graphicx} 
\usepackage{natbib}  
\usepackage{caption} 
\DeclareCaptionStyle{ruled}{labelfont=normalfont,labelsep=colon,strut=off} 
\usepackage{algorithm}
\usepackage{algorithmic}
\usepackage{graphicx}
\usepackage{epstopdf}
\usepackage{booktabs}
\usepackage{subfigure}
\usepackage{colortbl}
\definecolor{mygray}{gray}{.8}
\definecolor{mypink}{rgb}{.99,.91,.95}
\definecolor{mycyan}{cmyk}{.3,0,0,0}
\usepackage{amsfonts}
\usepackage{algorithmic}
\usepackage{algorithm}
\usepackage{float}
\usepackage{array}
\usepackage{adjustbox}
\usepackage{hyperref}
\makeatletter
\newcommand{\removelatexerror}{\let\@latex@error\@gobble}
\makeatother
\usepackage{amsmath}
\usepackage{latexsym}
\usepackage{multirow}
\usepackage{multicol}
\usepackage{balance}
\usepackage{newfloat}
\usepackage{listings}

\floatstyle{ruled}
\newfloat{listing}{tb}{lst}{}
\floatname{listing}{Listing}

    \title{EEGDnet: Fusing Non-Local and Local Self-Similarity for \\ 1-D EEG Signal Denoising with 2-D Transformer }

\affiliations{
   
    Peng Yi \textsuperscript{\rm 1} \equalcontrib,
    Kecheng Chen \textsuperscript{\rm 1} \equalcontrib,
     Zhaoqi Ma\textsuperscript{\rm 1},
     Di Zhao \textsuperscript{\rm 2},
     Xiaorong Pu \textsuperscript{\rm 1} \footnote{Corresponding author (puxiaor@uestc.edu.cn).},
     Yazhou Ren \textsuperscript{\rm 1}, \\
    
     \textsuperscript{\rm 1}School of Computer Science and Engineering, \\University of Electronic Science and
Technology of China, Chengdu, 611731, China\\
\textsuperscript{\rm 2} Institute of Computing Technology, \\Chinese Academy of Sciences, Beijing, 100080, China
    
}

\usepackage{bibentry}
\begin{document}

\maketitle

\begin{abstract}
Electroencephalogram (EEG) has shown a useful approach to produce a brain-computer interface (BCI). One-dimensional (1-D) EEG signal is yet easily  disturbed by certain artifacts (a.k.a. noise) due to the high temporal resolution. Thus, it is crucial to remove the noise in received EEG signal. 
Recently, deep learning-based EEG signal denoising approaches have achieved impressive performance compared with traditional ones. 
It is well known that the characteristics of self-similarity (including \textit{non-local} and \textit{local} ones) of data (e.g., natural images and time-domain signals) are widely leveraged for denoising. However, existing deep learning-based EEG signal denoising methods ignore either the non-local self-similarity (e.g., 1-D convolutional neural network) or local one (e.g., fully connected network and recurrent neural network).
To address this issue,  we propose a novel 1-D EEG signal denoising network with 2-D transformer, namely EEGDnet. Specifically, we comprehensively take into account the non-local and local self-similarity of EEG signal through the transformer module.  By fusing non-local self-similarity in self-attention blocks and local self-similarity in feed forward blocks, the negative impact caused by noises and outliers can be reduced significantly.
Extensive experiments show that, compared with other state-of-the-art models, EEGDnet achieves much better performance in terms of both quantitative and qualitative metrics.

\end{abstract}


\maketitle

\section{Introduction}

Brain-computer interface (BCI) provides a communication channel between the brain and external devices, which is widely used in bio-engineering and neuroscience \cite{birbaumer2006physiological}. 
According to the mechanism of signal acquisition, BCI can be divided into two streams, i.e., intrusive BCI and non-intrusive BCI. The intrusive BCI is prone to cause the patient's immune response as the electrodes must be inserted into the patient's brain, leading to the decline or even disappearance of the signal quality. Researchers thus prefer to use non-invasive BCI due to its lower cost and higher security.

Electroencephalogram (EEG) is a dominant non-invasive BCI \cite{minguillon2017trends,qin2018high}. As result of the high temporal resolution, EEG can be easily disturbed by complex noises \cite{jiang2019removal}, such as cardiac artifact, ocular artifact, muscle artifact as well as external noise. The effective denoising of EEG signal thus is an urgent and primary step for AI-driven EEG signal understanding. 
Among a variety of noises, external noise can be removed by reasonable tuning \cite{cai2020feature}, including replacing electrodes and using 50Hz (power frequency) traps. In addition, through setting reference channel, removing cardiac artifacts is a convenient operation. 

Based on above-mentioned analysis, EEG denoising task would mainly take account into the removals of ocular artifact and muscle artifact.
Ocular artifact usually generates obvious artifacts in EEG signal. The origin of ocular artifact is eye movement and blinks,  which can propagate over the scalp and be recorded by EEG activity. Muscle artifact can be caused by any muscle proximity to signal recording sites contraction and stretch, the subject talks, sniffs, swallows, etc.

For EEG signal denoising approaches, quite a few traditional and deep learning-based methods have been developed.
Traditional approaches have two streams: 1) Regression and filtering-based approaches (e.g., adaptive filtering-based approaches \cite{he2006removal}). 2) Decomposing EEG data and noise data into other domains (e.g., wavelet transform approach \cite{2001EEG}, blind source separation approach \cite{klados2011reg}).
Nevertheless, traditional approaches suffer from some drawbacks. For example, regression-based approaches only work if reference channels can be available. Filtering-based methods may eliminate useful EEG signals during artifact deletion. Meanwhile, decomposing EEG data into other domains are computationally complex and time-consuming, which is not suitable for fast response applications in real-world.

Thanks to the improvement of computing power and large amount of EEG data, researchers tend to utilize deep learning-based approaches to perform noise suppression of EEG signal in recent years. 
Generally, feedforward neural network (FNN) \cite{bebis1994feed}, convolutional neural network (CNN) \cite{albawi2017understanding} or recurrent neural network (RNN) \cite{zaremba2014recurrent} are leveraged to construct deep learning-based EEG signal denoising models. However, existing deep learning-based EEG signal denoising approaches ignore either the non-local self-similarity or local one, although their performance is improved significantly compared with traditional methods.

To address those issues, we propose to fuse the non-local and local 
self-similarity of EEG signal together, which is the intrinsic motivation in this paper. By doing so, we propose a novel EEG signal denoising network with 
transformer to perform the noise removal of EEG signal, namely EEGDnet.
Specifically, we firstly propose to leverage the 2-D transformer to comprehensively learn  non-local self-similarity in self-attention blocks and local self-similarity in feed forward blocks. Then, we propose to stack multiple transformer modules to explore an optimal denoising pattern with end-to-end fashion. Finally, we carefully balance resource consumption and denoising performance to cater the low-power desires of embedded BCI device.

The contributions of this paper can be summarized into three-folds:
\begin{itemize}
    \item Compared with existing deep learning-based
    EEG signal denoising methods, EEGDnet is the first attempt to comprehensively take into account the non-local and local self-similarity of EEG signal.
    \item EEGDnet is an easy yet effective model to introduce the 2-D transformer into 1-D time-domain signal denoising tasks.
    \item EEGDnet enjoys the advantage of lower model parameters and computation consumption on embedded BCI device, which have more competitive edge on real world.
\end{itemize}


\section{Related Work}
\label{sec related}

\subsection{Traditional EEG signal denoising approaches}

The simplest and most commonly used approach is the regression-based approach \cite{klados2011reg}. It is applied under the assumption that each channel is the cumulative sum of pure EEG data and a proportion of artifact \cite{sweeney2012artifact}, using exogenous reference channels (i.e., electrooculogram (EOG), electrocardiogram (ECG) ) to omit different artifacts. However, the regression-based approach cannot handle electromyogram (EMG) signals due to the absence of an exogenous reference channel.

Wavelet transform approach \cite{2001EEG}, transforming a time domain signal into time and frequency domain, has good time-frequency features relative to Fourier transform due to the better tunable time-frequency tradeoff and superiority of non-stationary signal analysis.

The Blind Source Separation (BSS) approach \cite{klados2011reg,1996Independent,2000EEG,2013The,2014cca} decomposes the EEG signal into components, distributes them to the neural source and the artifactual source and reconstructs a clean signal by recombining the neural components. However, BSS approach can only be used when a large number of electrodes are available.

Numerous filtering approaches are employed in the cancelation of artifacts from the EEG, for instance, adaptive filtering, wiener filtering and Bayes filtering, in which different approaches implemented with different principle of optimization \cite{he2006removal}. The two main types of filters are the adaptive filtering \cite{marque2005adaptive}, which requires an additional reference channel, and the wiener filtering \cite{somers2018generic}, which requires calibration.

\begin{figure*}[!h]
\centering
\includegraphics[scale=1]{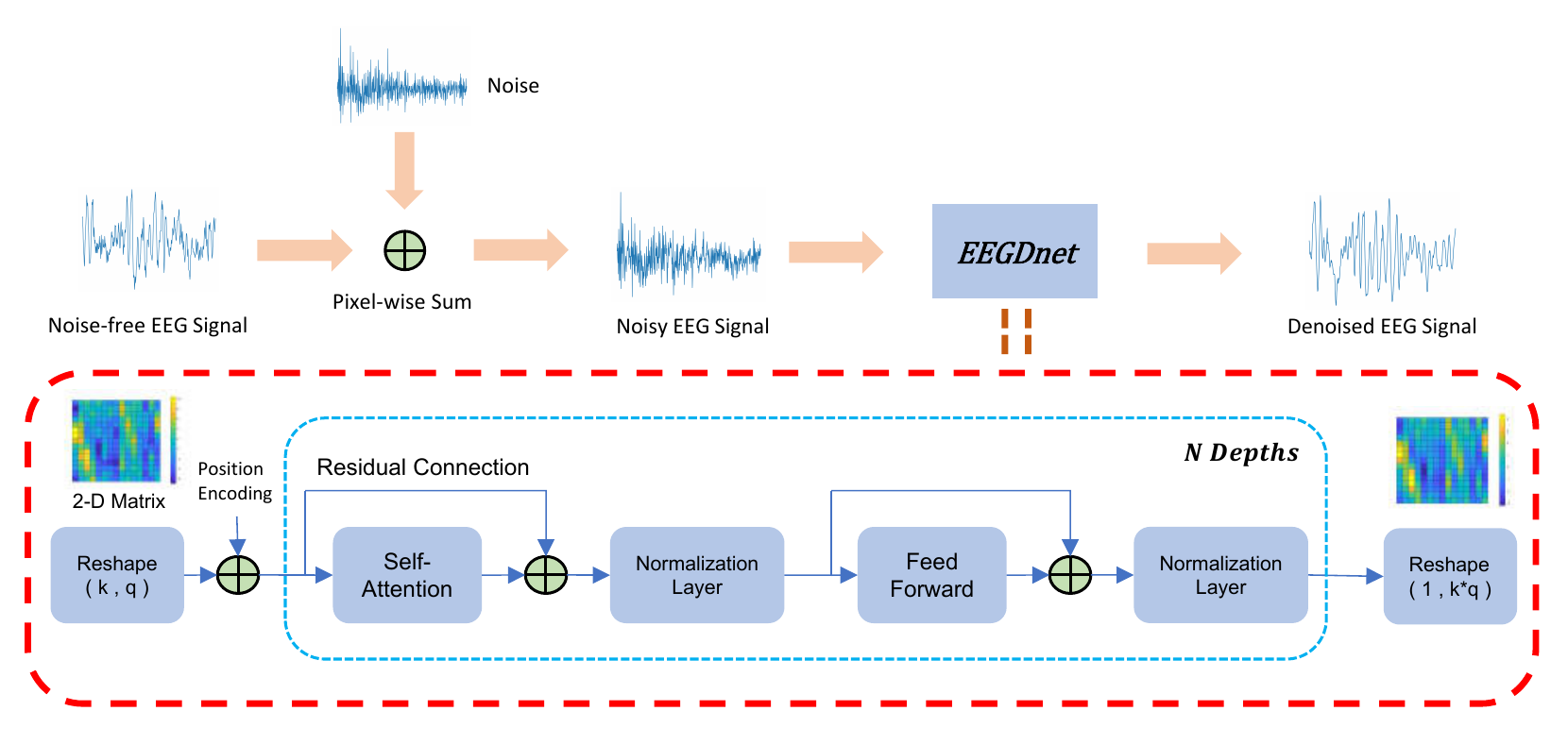} 
\caption{The architecture of the training phase of the EEG denoising model based on the EEGDnet model. 
First, the noise-free EEG signal and the artifact (ocular artifact or muscle artifact) are pixel-wise summed according to a certain signal-to-noise ratio.
The result, namely noisy EEG signal, and corresponding noise-free one form the input and the label of EEGDnet, respectively.
Inside the EEGDnet, the signal is transformed into a 2-D matrix through the reshape layer at the beginning. The opposition is true at the end of the network.
And then sequentially through the self-attention block, normalization layer, feed forward block and normalization layer.
Note that there are residual connections in the network. $N Depths$ indicates the number of self-attention blocks as well as other components.
Finally, these features that can distinguish EEG signal and noise are automatically reconstructed by minimizing the objective function (MSE) to get the denoised EEG signal.}
\label{fig EEGDnet}
\end{figure*}

\subsection{Deep learning-based EEG signal denoising approaches}

Deep learning has been widely used in computer vision \cite{krizhevsky2012imagenet, simonyan2014very, szegedy2016rethinking, he2016deep} and natural language processing \cite{mikolov2013efficient, vaswani2017attention, sutskever2014sequence}, and has achieved remarkable success in the last years. Furthermore, deep learning has been applied to other fields, achieving comparable results with traditional methods including EEG denoising \cite{yang2018automatic, hanrahan2019noise, aynali2020noise, sun2020novel}.
With the development of computing power and data volume, deep learning can learn essential characteristics of neural oscillations hidden in the data and eliminate artifact from other parts of biology.

Yang $et~al.$ \cite{yang2018automatic} proposed a fully connected neural network (namely DLN) to remove ocular artifact. The structure of DLN is relatively simple containing only 3 hidden layers and sigmoid activation function. Compared to traditional approaches, DLN is time-saving and do not need manual involvement when reducing noise.

Based on convolutional neural network (CNN), Sun $et~al.$ \cite{sun2020novel} presented a one-dimensional residual CNN (1D-ResCNN) model, achieving better performance in some cases compared with traditional methods. In order to better reduce noise, 1D-ResCNN leverages convolution kernels of different scales to process data in parallel. Meanwhile, residual blocks are used to prevent vanishing gradient problem.

Zhang $et~al.$ \cite{zhang2020eegdenoisenet} summarizes the previous work and proposes four basic models both for muscle artifact and ocular artifact. The four basic models are based on FNN, CNN, CNN with residual blocks and RNN, respectively.

\subsection{Transformer}
In the very recent years, transformer \cite{vaswani2017attention}, a type of self-attention-based \cite{hao2020self} neural networks originally applied for natural language processing tasks \cite{DBLP:journals/corr/abs-2106-03106}, has attracted the attention of most researchers in AI communities. As it develops, it has almost replaced RNN with its excellent effects, such as exploring long-range relationship, better scalability to high-complexity models and so on. We can observe that transformer has been applied to image classification \cite{DBLP:conf/cvpr/LanchantinWOQ21, alexey2020image}, object detection \cite{DBLP:conf/aaai/YangWCLQ20, DBLP:conf/iclr/ZhuSLLWD21, DBLP:conf/eccv/CarionMSUKZ20} and even natural image restoration \cite{DBLP:journals/corr/abs-2106-03106, DBLP:journals/corr/abs-2108-10257}.

\section{Proposed approach}
\subsection{The overall framework of proposed EEGDnet}
\label{sec method}
In this paper, we propose a EEG signal denoising model to carry out the one-dimensional (1-D) signal noise reduction task.
Note that 1) The proposed model aims to well fuse the 
non-local and local self-similarity of 1-D EEG signal into a 2-D transformer module. 2) The proposed model can also be extended to other types of 1-D signals.

The overall framework of our approach is shown in Figure \ref{fig EEGDnet}. It can be seen as an end-to-end model, i.e., giving a noisy EEG signal as the input and outputting the corresponding noise-free one.
The degradation process of acquired EEG signal can be represented as:
\begin{equation}
    \label{eq mixed}
    y=x+\lambda n
\end{equation}
where $y\in \mathbb{R}^{1\times N}$ ($N$ is the number of signal point in a period) denotes an acquired noisy EEG signal, $x\in \mathbb{R}^{1\times N}$ denotes a theoretical noise-free EEG signal, $\lambda$ and $n\in \mathbb{R}^{1\times N}$ denote the relative contribution of artifacts and artifact (ocular artifact or muscle artifact), respectively.

For deep learning-based EEG signal denoising approaches, the denoising procedure can usually be regarded as a nonlinear mapping function from noisy EEG signal $y$ to the corresponding estimated noise-free one $\hat{x}$, which can be formulated as:
\begin{equation}
    \hat{x} = F(y;\theta),
\end{equation}
where $F(\cdot)$ denotes a nonlinear mapping function, specifically a neural network. $\theta$ is the learnable parameters of the neural network.
For objective, following widely-adopted image/signal denoising methods \cite{marmolin1986subjective}, we leverage Mean Square Error (MSE) as the loss function to minimize the difference between estimated the noise-free EEG signal $\hat{x}$ and corresponding noise-free one $x$, denoted as $\mathcal{L}(x,\hat{x})$.
The overall EEG signal denoising algorithm is presented in Algorithm \ref{alg DeT}.

\begin{figure}[!ht]
    \renewcommand{\algorithmicrequire}{\textbf{Input:}}
	\renewcommand{\algorithmicensure}{\textbf{Output:}}
	\removelatexerror
	\begin{algorithm}[H]
		\caption{EEGDnet}
		\label{alg DeT}
		\begin{algorithmic}[1]
			\REQUIRE 
			$X_{s} = \{ x_{s}^{i}\}_{i=1,\dots,n}$,  $Y_{s} = \{ y_{s}^{i}\}_{i=1,\dots,n} $ denote ground-truth and noisy signal in training dataset, respectively; $X_{t} = \{ x_{t}^{i}\}_{i=1,\dots,m}$,  $Y_{t} = \{ y_{t}^{i}\}_{i=1,\dots,m} $ denote ground-truth and noisy signal in validation dataset, respectively;  
			\ENSURE $F(;\theta)$ - a well trained transformer model
			\STATE {Initialize ${\theta}$ with random values}
			\REPEAT
			    \STATE Obtain the estimated noise-free output $\hat{x}_s$ by $F(y_s;\theta)$ on training set;
			    \STATE Calculate the difference between estimation and ground-truth on training set by $\mathcal{L}(x_s,\hat{x}_s)$;
			    \STATE Update ${\theta}$ using Adam optimizer;
			    \STATE Obtain the estimated noise-free output $\hat{x}_t$ by $F(y_t;\theta)$ on validation set;
			    \STATE Calculate the difference between estimation and ground-truth on validation set $\mathcal{L}(x_t,\hat{x}_t)$;
			\UNTIL  $\mathcal{L}(x_t,\hat{x}_t)$ converges;
			\RETURN $F(;{\theta})$
		\end{algorithmic}
	\end{algorithm}
\end{figure}

\subsection{Transformer for EEG Signal Denoising}
In this paper, a novel EEG signal denoising model (namely EEGDnet) with transformer is proposed. Specifically, we follow original 2-D transformer \cite{vaswani2017attention} and incorporate it into the denoising scheme. It should be noted that an advantage of this setup is that scalable natural language processing transformer architectures and their efficient implementations can be leveraged as much as possible. 

The detailed transformer architecture in  proposed EEGDnet model (as shown in red dashed box of Figure \ref{fig EEGDnet}) consists of four parts including \textbf{reshape layer}, \textbf{self-attention block}, \textbf{feed forward block} and \textbf{normalization layer}. 
Specifically, the 1-D EEG signal is reshaped into 2-D form firstly and then it was sent into the 2-D transformer encoder, which consists of alternating layers of self-attention blocks and feed forward blocks. Normalization layer and residual connections are applied after every block. The output of transformer encoder is reshaped into one dimension as the final output. These modules will be described.

\subsubsection*{Reshape layer:}
In computer vision fields, an 2-D image can be split into a variety of patches in order to leverage the non-local self-similarity. Inspired by this \cite{ alexey2020image}, we propose to split an epoch of EEG signal into several segments. In other words, an EEG signal $x \in \mathbb{R}^{1 \times N}$ is split into fixed-size segments $s \in \mathbb{R}^{k \times q}$. Note that $N$ is the length of an epoch, $k$ denotes the input sequence length and $q$ denotes the dimensions of each segment. Correspondingly, at the end of the EEGDnet model, the data will be reverted to original one-dimensional form.
Standard learnable 1D position embeddings are then added to the segments embeddings to retain positional information. The resulting sequence of embedding vectors serves as input to the transformer encoder.


\subsubsection*{Self-attention blocks:}
In self-attention blocks, each segment of the signal corresponds to a learnable query and a set of learnable key values.
An attention function can be described as mapping a query and a set of key-value pairs to an output, where the query, keys, values, and output are all vectors. The output is computed as a weighted sum of the values, where the weight assigned to each value is computed by a compatibility function of the query with the corresponding key.
In simple terms, the attention function allows finding the weight of any segment on any segment, thus exploiting the non-local self-similarity between segments.

Furthermore, to enhance the ability of self-attention blocks, multi-head attention \cite{vaswani2017attention} are introduced as an expansion and can be artificially set. When the number of heads, noted as $N Heads$, equals to one, multi-head attention degenerates into a single attention head.

\subsubsection*{Feed forward blocks:}
To take full advantage of the local self-similarity, feed forward blocks use fully connected layers and nonlinear activation function to process each segment.
That is, from an epoch perspective, the feed forward blocks exploit local self-similarity within each segment.
Besides, dropout regularization \cite{srivastava2014dropout} is utilized to enhance the robustness of EEGDnet model.

It should also be noted that different from original transformer, we replace Rectifier Linear Unit (ReLU) \cite{2011Deep} with Parametric Rectified Linear Unit (PReLU) \cite{he2015delving} in feed forward blocks. Unlike words or image embedding, EEG signal contains both positive and negative values. Consequently, feeding values that are beyond the usual range of features can cause large gradients to back propagate \cite{haldar2019applying}. This can permanently shut activation functions like ReLU due to vanishing gradients \cite{clevert2015fast}, which is unacceptable to the network. 

Compared with ReLU, PReLU has one more parameter $\textit{a}$, which represents the derivative if the input is less than zero. This ensures that PReLU has a non-zero output regardless of the input, easing the problem of neuronal death. 


\subsubsection*{Normalization layer and residual connection:}
Training state-of-the-art deep neural networks is computationally expensive. One
way to reduce the training time is to normalize the activities of the neurons. Generally speaking, batch normalization \cite{ioffe2015batch} is adopted for image processing while layer normalization \cite{ba2016layer} for natural language processing. Given the 1-D nature of the EEG signal, the EEGDnet model uses layer normalization as the normalization layer.

Vanishing gradients and exploding gradients are commonly encountered problem in the field of deep learning. The EEGDnet model alleviates this problem using a simple residual structure \cite{he2016deep}.






\begin{table*}
\caption{Average performances of all SNRs (from -7dB to 2dB). The smaller ${RRMSE}_{temporal}$ and ${RRMSE}_{spectral}$, and the larger $CC$, the better denoising effect. 
The baseline of EEGDnet consists of $6~Depths$ and $1~Head$ with $k ~\times ~q$ equals to $8~\times~64$.
Note that all the models are trained and tested on the same data set. For ${RRMSE}_{temporal}$, ${RRMSE}_{spectral}$, the lower the better. For $CC$,
the higher the better. The best result is shown in bold.}
\begin{adjustbox}{max width= \textwidth}
\renewcommand{\arraystretch}{1.4}
\label{table model}
\centering
\begin{tabular}{c|cc|ccc|ccc}
\toprule
\multirow{2}*{Model}  & \multicolumn{2}{c}{ Self-Similarity } &\multicolumn{3}{c}{ Ocular Artifact }   &  \multicolumn{3}{c}{ Muscle Artifact} \\
&  local & non-local & ${RRMSE}_{temporal}$ & ${RRMSE}_{spectral}$ & $CC$  &  ${RRMSE}_{temporal}$ & ${RRMSE}_{spectral}$ & $CC$ \\
\midrule
DLN \cite{yang2018automatic} & & \checkmark & 0.699 & 0.579& 0.720 & 0.917 & 1.081 & 0.609  \\
SCNN \cite{zhang2020eegdenoisenet} & \checkmark &   & 0.620 & 0.526 & 0.791 & 0.750 & 0.697 & 0.706 \\
1D-ResCNN \cite{sun2020novel} & \checkmark & & 0.630 & 0.588 & 0.776 & 0.746 & 0.680 & 0.692 \\
RNN \cite{zhang2020eegdenoisenet} &  & \checkmark & 0.740 & 0.696 & 0.677 & 0.785 & 0.775 & 0.636 \\
\rowcolor{mygray}
EEGDnet & \checkmark & \checkmark & \textbf{0.497} & \textbf{0.491} & \textbf{0.868} & \textbf{0.677} & \textbf{0.626} & \textbf{0.732} \\
\bottomrule
\end{tabular}
\end{adjustbox}
\end{table*}

\section{Experiments and results}
\label{sec experiments}
\subsection{Experimental Setup}

\subsubsection*{Data set:} 
To evaluate the effectiveness of the proposed EEGDnet model, a widely-adopted EEG signal data set \cite{zhang2020eegdenoisenet} is used in this paper. Specifically, this data set consists of 4514 EEG epochs, 3400 ocular artifact epochs and 5598 muscle artifact epochs. The sampling time of each epoch is 2 seconds with 256 sampling rate. During the training, simulated mixed signals can be generated according to Eq. \ref{eq mixed} with a uniformly distributed signal-to-noise ratio (SNR) from -7dB to 2dB.
To enhance the diversity of data, EEG epochs are randomly combined with ocular artifact epochs and muscle artifact epochs for ten times, respectively. 

\subsubsection*{Training details:}
The deep learning framework (PyTorch) is applied to construct our EEGDnet model. The entire model is optimized by Adam \cite{kingma2014adam} optimizer for 10K epochs with a learning rate of $5e^{-5}$ and betas of $(0.5,0.9)$. The batch size is 1K.

\subsubsection*{Comparing methods:}
We compare the proposed EEGDnet model with four existing state-of-the-art deep learning-based EEG denoising methods:
\begin{itemize}
  \item 
  Deep learning network (DLN) \cite{yang2018automatic}: A fully connected neural network with three hidden layers and sigmoid function.
  \item
  Recurrent neural network (RNN) \cite{zhang2020eegdenoisenet}: The network contains in order a Long Short-Term Memory \cite{hochreiter1997long} network and three fully connected layers with the ReLU activation function and dropout regularization.
  \item
  Simple convolutional neural networks (SCNN) \cite{zhang2020eegdenoisenet}: The network contains four 1D-convolution layers with small $1*3$ kernels, $1$ stride, and $64$ feature maps. Each 1D-convolution layer was followed by a batch normalization layer and a ReLU activation function. At the end of the network, there is a fully connected layer to reconstruct the signal.
  \item
  One-dimensional residual convolutional neural networks (1D-ResCNN) \cite{sun2020novel}: The structure of the network mainly has three parallel sub-modules which have different convolutional kernels and residual connections. Each 1-D convolution layer was followed by a batch normalization layer and a ReLU activation function. Data from three sub-modules are merged. 
\end{itemize}






\subsubsection*{Evaluation measures:}
In terms of differences and correlations both in time and frequency domains, three quantitative metrics \cite{zhang2020eegdenoisenet} are applied to evaluate the performance of different models, including Relative Root Mean Square Error in the temporal domain (${RRMSE}_{temporal}$), Relative Root Mean Square Error in the spectral domain (${RRMSE}_{spectral}$) and the correlation coefficient ($CC$). 

With the development of the EEG equipment towards miniaturization \cite{minguillon2017trends}, the computational and parametric quantities of the model must be considered due to the limited computational and storage resources. Therefore, we utilize the amount of calculation(${FLOPs}$), number of parameters(${Param}$) and storage size to reflect the actual usability of the models.




\subsection{Results}

\begin{table}
\caption{The amount of computation, number of parameters and storage size of models.
The baseline of EEGDnet consists of $6~Depths$ and $1~Head$ with $k ~\times ~q$ equals to $8~\times~64$. For flops, parameters and storage size, the lower the better.  The best result is shown in bold. }
\begin{adjustbox}{max width=\columnwidth}
 \renewcommand{\arraystretch}{1.7}

\label{flops and parameters}
\centering
\begin{tabular}{cccc}
\toprule
Model & $FLOPs(M)$ & $Param(M)$ & $StorageSize(MB)$\\
\midrule
DLN \cite{yang2018automatic} & 1.05 & 1.05 & 64.1\\
SCNN \cite{zhang2020eegdenoisenet} & 36.14 & 16.81 & 1026.3\\
1D-ResCNN \cite{sun2020novel} & 42.33 & 8.42 & 514.2\\
RNN \cite{zhang2020eegdenoisenet} & \textbf{0.79} & 0.53 & 32.1\\
\rowcolor{mygray}
EEGDnet & 1.44 & \textbf{0.18} & \textbf{11.1}\\
\bottomrule
\end{tabular}
\end{adjustbox}
\end{table}

\subsubsection{Quantitative Results}
Table \ref{table model} shows the quantitative performance of denoising results for both ocular and muscle artifacts. Some observations can be obtained as following.

\begin{itemize}
    \item From the perspective of DLN and RNN, they can only leverage non-local self-similarity resulting worst performance in denoising.
    \item Stacking of convolutional layers in both SCNN and 1D-ResCNN allows network have large receptive field, which means non-local self-similarity can be exploited to some extent implicitly.
    \item By fusing non-local self-similarity and local one explicitly, EEGDnet can have better performance.
\end{itemize}

\begin{table*}[!h]
\renewcommand{\arraystretch}{1.0}
\caption{The effect of $k ~ \times ~ q$ on the noise reduction performance of EEGDnet. Note $k ~ \times ~ q$ must be equal to $N=512$ with $1~Head$ and $6~Depths$.}
\begin{adjustbox}{max width=\textwidth}
\label{table kq}
\centering
\begin{tabular}{c|c|ccc|ccc}
\toprule
\multirow{2}{*}{$k ~ \times ~ q$} & \multirow{2}{*}{Params(K)} & \multicolumn{3}{c}{ Ocular artifact }   &  \multicolumn{3}{c}{ Muscle artifact} \\
 & & ${RRMSE}_{temporal}$ & ${RRMSE}_{spectral}$ & $CC$  &  ${RRMSE}_{temporal}$ & ${RRMSE}_{spectral}$ & $CC$ \\
\midrule
$2~ \times ~256$  & 2889 & 0.636 & 0.568 & 0.769 & 0.727 & 0.674 & 0.690 \\
$4~ \times ~128$  & 724  & 0.554 & 0.529 & 0.832 & 0.698 & 0.648 & 0.715 \\
$8~ \times ~64$   & 182  & 0.497 & 0.491 & 0.868 & 0.677 & 0.626 & 0.732 \\
$16~ \times ~32$  & 46   & 0.469 & 0.476 & 0.882 & 0.652 & 0.600 & 0.749 \\
$32~ \times ~16$  & 12   & 0.459 & 0.476 & 0.885 & 0.642 & 0.597 & 0.743 \\
$128~ \times ~4$  & 1.3  & 0.691 & 0.770 & 0.711 & 0.776 & 0.804 & 0.646 \\

\bottomrule
\end{tabular}
\end{adjustbox}
\end{table*}

\begin{table*}[!h]
\renewcommand{\arraystretch}{1.0}
\caption{The effect of the depths of EEGDnet on the noise reduction performance. Note that $k ~ \times ~ q$ and $N~Heads$ equal to $8~ \times ~64$ and $1~Head$, respectively.}
\begin{adjustbox}{max width=\textwidth}
\label{table depths}
\centering
\begin{tabular}{c|c|ccc|ccc}
\toprule
\multirow{2}{*}{$N~Depths$} & \multirow{2}{*}{Params(K)} & \multicolumn{3}{c}{ Ocular artifact }   &  \multicolumn{3}{c}{ Muscle artifact} \\
 & & ${RRMSE}_{temporal}$ & ${RRMSE}_{spectral}$ & $CC$  &  ${RRMSE}_{temporal}$ & ${RRMSE}_{spectral}$ & $CC$ \\
\midrule
$2$   & 67  & 0.492 & 0.509 & 0.872 & 0.663 & 0.608 & 0.742 \\
$4$   & 124 & 0.487 & 0.491 & 0.873 & 0.669 & 0.620 & 0.737 \\
$6$   & 182 & 0.497 & 0.491 & 0.868 & 0.677 & 0.626 & 0.732 \\
$8$   & 240 & 0.512 & 0.490 & 0.859 & 0.680 & 0.639 & 0.729 \\
$10$  & 298 & 0.526 & 0.495 & 0.850 & 0.681 & 0.633 & 0.727 \\

\bottomrule
\end{tabular}
\end{adjustbox}
\end{table*}

\begin{table*}[!h]
\caption{The effect of number of heads, denoted as $N~Heads$, in self-attention blocks on the noise reduction performance. Note that $k ~ \times ~ q$ and $N~Depths$ equal to $8~ \times ~64$ and $6~Depths$, respectively.}
\renewcommand{\arraystretch}{1.0}
\begin{adjustbox}{max width=\textwidth}

\label{table head}
\centering
\begin{tabular}{c|c|ccc|ccc}
\toprule
\multirow{2}{*}{$N~Heads$} & \multirow{2}{*}{Params(K)} & \multicolumn{3}{c}{ Ocular artifact }   &  \multicolumn{3}{c}{ Muscle artifact} \\
 & & ${RRMSE}_{temporal}$ & ${RRMSE}_{spectral}$ & $CC$  &  ${RRMSE}_{temporal}$ & ${RRMSE}_{spectral}$ & $CC$ \\
\midrule
$1$   & 182  & 0.497 & 0.491 & 0.868 & 0.677 & 0.626 & 0.732 \\
$2$   & 305  & 0.501 & 0.492 & 0.866 & 0.678 & 0.636 & 0.731 \\
$4$   & 502  & 0.526 & 0.508 & 0.850 & 0.687 & 0.651 & 0.724 \\
$8$   & 895  & 0.562 & 0.526 & 0.828 & 0.695 & 0.659 & 0.716 \\
$16$  & 1682 & 0.577 & 0.549 & 0.820 & 0.696 & 0.654 & 0.715 \\

\bottomrule
\end{tabular}
\end{adjustbox}
\end{table*}

Meanwhile, there are strict requirements on the amount of operations and the number of parameters of the model, since EEG noise reduction is usually applied to microchips on brain-computer interface devices. It is desirable for the model to have a low number of model parameters as well as computational consumption. Table \ref{flops and parameters} illustrates  the amount of computation, number of parameters and storage size of models. We can notice that EEGDnet model, compared with other deep learning-based denoising methods, not only has a smaller number of parameters and storage size but also achieves state-of-the-art denoising effects. The proposed EEGDnet will be a better choice in real-world application.



\begin{figure*}[!h]
\centering
\subfigure[]{
\begin{minipage}[b]{.9\linewidth}
\centering
\includegraphics[width=0.90\textwidth]{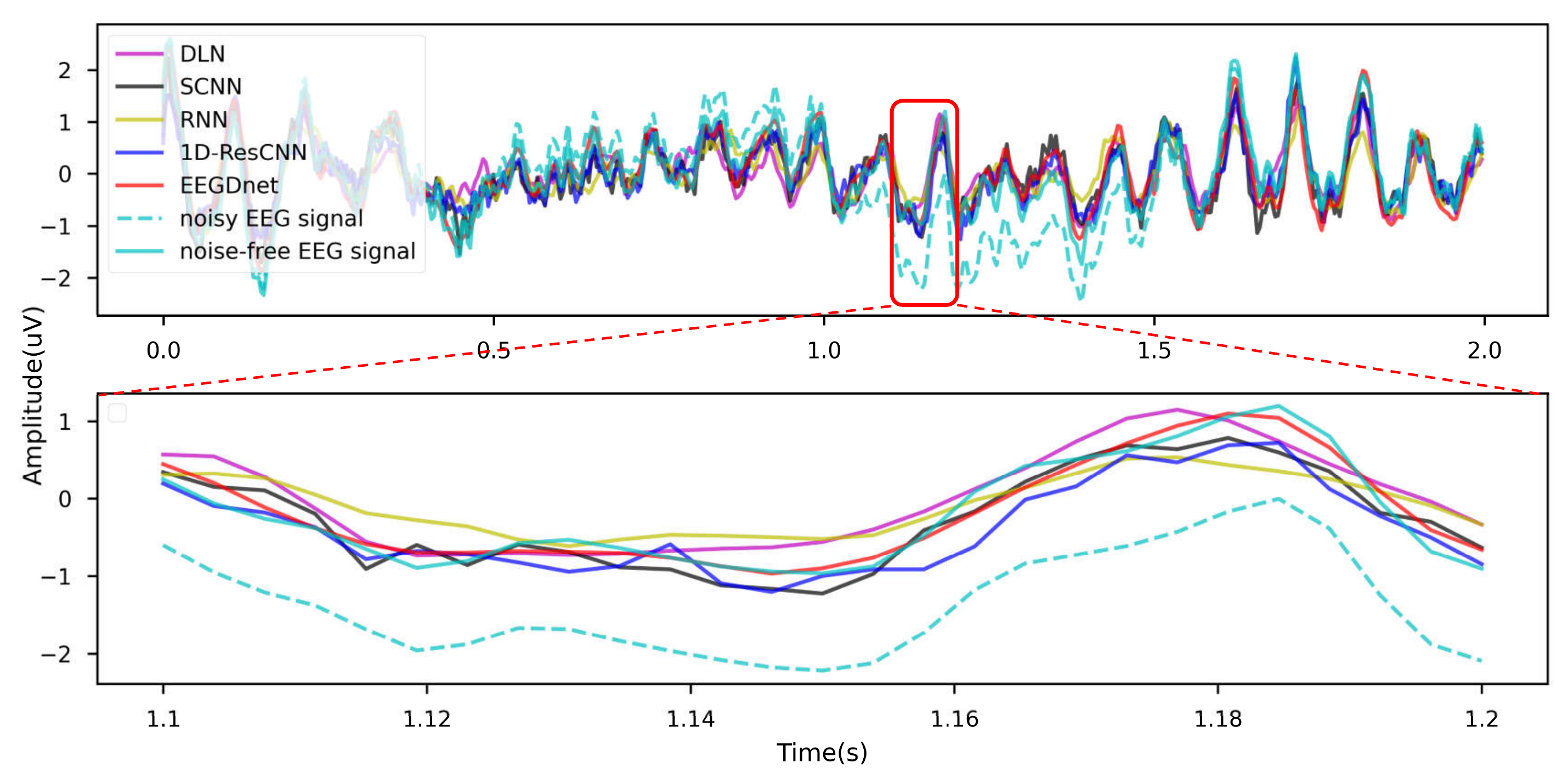} 
\end{minipage}
}
\subfigure[]{
\begin{minipage}[b]{.9\linewidth}
\centering
\includegraphics[width=0.90\textwidth]{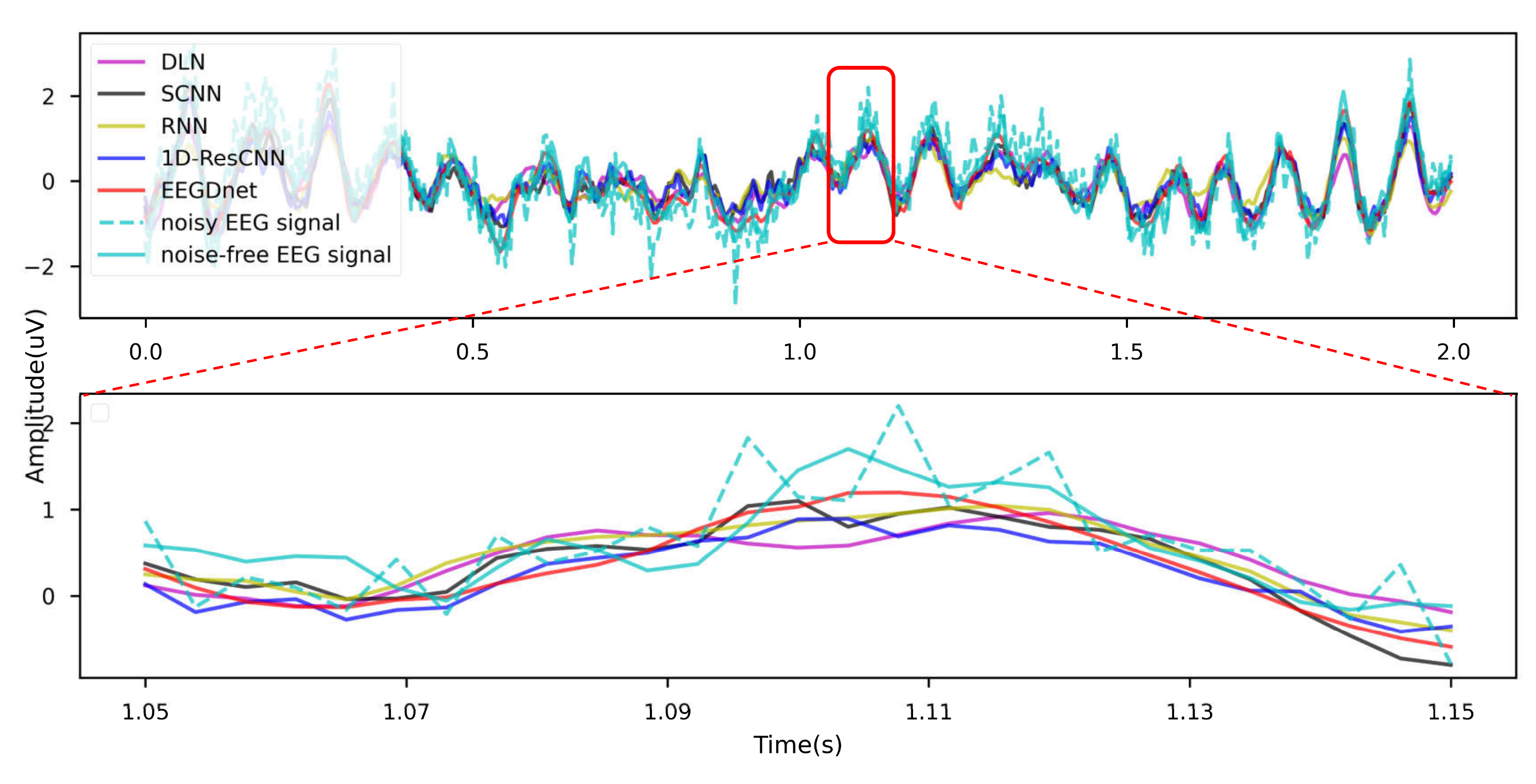} 
\end{minipage}

}

\caption{Visualized examples of denoising results with different state-of-the-art methods. a) Denoising on Ocular Artifact b) Denoising on Muscle Artifact. For each sub-figure, an easily-observed window (red box and corresponding zoomed-in view) is selected for better comparison. Note that the amplitude is normalized and the time domain sampling rate is 256 SPS. Regarding two examples, we can observe that the denoising results of our proposed EEGDnet are closer to the ground-truth signals. Please zoom in for better view. }
\label{figure_Qualitative}
\end{figure*}

\subsubsection{Qualitative Results}
The time-domain order waveforms image can reflect the EEG characteristics with the fluctuation of the waveform.
Figure \ref{figure_Qualitative} shows the qualitative results for noisy EEG signal. We have the following observations.
\begin{itemize}
    \item All the methods have certain denoising effects on the noisy EEG signal, while remarkable differences can be noticed in detail. 
    \item Notice that both DLN and RNN differ significantly in detail from the noise-free EEG signal. This may result from their inability to effectively exploit local self-similarity.
    \item SCNN and 1D-ResCNN are at a disadvantage in tracking the overall changes of the noise-free EEG signal, although they retain more local details. This is due to the almost fully convolutional network structure, which cannot exploit the non-local self-similarity.
    \item Overall, the EEGDnet model uses both non-local self-similarity to track overall changes in the signal and local self-similarity to retain detailed information.
\end{itemize}


\subsubsection{Ablation study}

As the first work to apply transformer to 1-D EEG signal noise reduction, ablation experiments are used to explored the effect of some hyperparameters on the noise reduction performance. 

We are interested in how the segmentation of different EEG signals (i.e., the shape of 2-D matrix) affects the noise reduction performance. According to the results in Table \ref{table kq}, we have the following observations:
\begin{itemize}
    \item As $q$ becomes smaller, the number of parameters is significantly reduced, mainly due to the reduction in the number of parameters required for the fully connected layer in the feed forward blocks.
    \item When $k$ is extremely small, self-attention blocks can not play its role. Also, when $q$ is extremely small, the segment is similar to a point. Both of the above cases lead to degradation of the network to a fully connected neural network which can not exploit local similarity.
    \item The noise reduction effect is best when $k$ and $q$ are close to each other, due to the ability to balance local self-similarity and non-local one well.
\end{itemize}

Table \ref{table depths} and Table \ref{table head} illustrate the effect of different depths and different numbers of heads on the noise reduction effect, respectively.
Both hyperparameters (i.e., $N~Depths$ and $N~Heads$), to a certain extent, can describe the complexity of EEGDnet model.
It can be found that as the complexity of the network decreases, the noise reduction effect becomes progressively better instead. This may be owing to the fact that the data set we use is relative small and the complex network tends to be overfitting.

\section{Conclusion}
\label{sec conclusion}
Ocular and muscle artifacts denoising is an acknowledged difficult task in EEG-based BCI.
In this paper, by fusing non-local self-similarity in self-attention blocks and local self-similarity in feed forward blocks, transformer is first imposed to EEG denoising task.
The proposed method can not only significantly reduce the model parameters, but also improve the overall performance for EEG denoising.
Extensive experiments have demonstrated that our model outperforms existing denoising methods both in quantitative and qualitative results.
It is worth noting that EEGDnet model is a general model that can be also applied to other 1-D signal processing.
In the future work, we will extend the proposed model to other applications such as EEG features extraction and classification.


\balance

\end{document}